\documentclass[superscriptaddress,aps,prd,nofootinbib,twocolumn,showpacs]{revtex4-1}
\usepackage{amsfonts}
\usepackage{amsmath}
\usepackage{amssymb}
\usepackage{graphicx}
\usepackage{mathrsfs}
\usepackage{hhline,color}
\usepackage{pifont,ulem}
\usepackage{lmodern}
\usepackage{epsfig}
\usepackage{enumitem} 
\usepackage{ulem}
\usepackage[utf8]{inputenc}
\usepackage{multirow}

\usepackage{braket}

\newcommand{\ave}[1]{{\langle{#1}\rangle}}

\newcommand{\expm}{e^{-\beta\,E^+_p}}
\newcommand{\expmm}{e^{-2\beta\,E^+_p}}
\newcommand{\expmmm}{e^{-3\beta\,E^+_p}}
\newcommand{\expp}{e^{-\beta \,E^-_p}}
\newcommand{\expppp}{e^{-3\beta\,E^-_p}}
\newcommand{\exppp}{e^{-2\beta\,E^-_p}}

\newcommand{\pc}[1]{\ensuremath{\left(#1\right)}}
\newcommand{\chav}[1]{\ensuremath{\left\{#1\right\}}}
\newcommand{\ev}[1]{\ensuremath{\left\langle #1\right\rangle}}

\def\beq{\begin{equation}}
\def\eeq{\end{equation}}
\def\beqa{\begin{eqnarray}}
\def\eeqa{\end{eqnarray}}
\def\ban{\begin{eqnarray*}}
\def\ean{\end{eqnarray*}}
\def\bi{\begin{itemize}}
\def\ei{\end{itemize}}

\newcommand{\Z}{\mathbb{Z}}

\begin{document}

\title{Presence of a critical endpoint in the QCD phase diagram from the 
net-baryon number fluctuations}

\author{M\'arcio Ferreira}
\email{mferreira@teor.fis.uc.pt}
\affiliation{CFisUC, Department of Physics,
University of Coimbra, P-3004 - 516  Coimbra, Portugal}
\author{Pedro Costa}
\email{pcosta@uc.pt}
\affiliation{CFisUC, Department of Physics,
University of Coimbra, P-3004 - 516  Coimbra, Portugal}
\author{Constan\c ca Provid\^encia}
\email{cp@fis.uc.pt}
\affiliation{CFisUC, Department of Physics,
University of Coimbra, P-3004 - 516  Coimbra, Portugal}

\date{\today}

\begin{abstract}
The net-baryon number fluctuations for three-flavor quark matter are computed 
within the Polyakov extended Nambu--Jona-Lasinio model.
Two models with vanishing and nonvanishing vector interactions are considered.
While the former predicts a critical end point (CEP) in the phase diagram, 
the latter predicts no CEP.
We show that the nonmonotonic behavior of the susceptibilities in the phase 
diagram is still present even in the absence of a CEP.  
Therefore, from the nonmonotonic behavior of the susceptibilities solely, one 
cannot assume the existence of a CEP. We analyze other possible properties that 
may distinguish the two scenarios, and determine the behavior of the net-baryon 
number fluctuations and the velocity of sound along several isentropes, with 
moderate and small values. It is shown that the value of the susceptibilities 
ratios and the velocity of sound at two or three isentropic lines could 
possibly allow to distinguish both scenarios, a phase diagram with or without 
CEP. Smoother behaviors of these quantities may indicate the nonexistence of a 
CEP. We also discuss the critical behavior of the strange sector.
\end{abstract}


\maketitle

\section{Introduction}

The quest for the QCD chiral critical end point (CEP) in the phase diagram, 
together with the nature of the phase transition between hadron matter and quark 
gluon plasma (QGP), are open questions that have attracted the attention of the 
physical community for some years \cite{Halasz:1998qr}. Remarkable theoretical 
and experimental efforts \cite{Brambilla:2014jmp} are being made to unveil the 
rich details of the QCD phase structures \cite{Gupta:2011wh}. Experimentally, 
one of the main goals of the heavy ion collision (HIC) program is the possible 
existence and location of the CEP on the QCD phase diagram, with great 
developments over the last years 
\cite{Abelev:2009bw,Aggarwal:2010cw,Adamczyk:2013dal,Aduszkiewicz:2015jna}. 

In relativistic HIC, the measurement of fluctuations of conserved 
quantities, such as baryon, electric charge, and strangeness number, play a 
major role in the experimental search for the CEP. Indeed, experimental 
measurements of cumulants of net-proton (proxy for net-baryon), net-charge, and 
net-kaon (proxy for net-strangeness) are expected to carry significant amounts 
of information on the medium created by the collision (for a review, see 
\cite{Friman:2011pf,Asakawa:2015ybt,Braun-Munzinger:2015hba,Luo:2017faz}).
Fluctuations are studied by measuring event-by-event fluctuations: a given 
observable is measured on an event-by-event basis and its fluctuations are 
studied for the ensemble of events \cite{Braun-Munzinger:2015hba}.

Particularly relevant are the cumulants of the net-baryon number because a 
second-order phase transition occurs at the CEP, resulting in divergences of
correlation lengths for a static system of infinite size. 
The cumulants of the baryon number thus diverge at the CEP 
\cite{Stephanov:1998dy,Stephanov:1999zu}.
The study of the kurtosis \cite{Stephanov:2011pb} and the skewness 
\cite{Asakawa:2009aj} for the net-baryon number fluctuation distributions is 
essential as they are related to higher-order cumulants that can be extracted 
from event-by-event fluctuations in HIC experiments. 
Once they are constituted by ratios of cumulants they are independent of the 
volume of the system.

The study of fluctuations of conserved charges (baryon number, electric charge, 
and strangeness) at finite temperature and density has been done by using the 
(2+1) flavor Nambu$-$Jona-Lasinio (NJL) model in 
\cite{Luo:2017faz,Chen:2015dra,Fan:2016ovc,Fan:2017kym}. 
By using the (2+1) Polyakov$-$Nambu$-$Jona-Lasinio (PNJL) model, these 
fluctuations were investigated at finite temperature in 
\cite{Fu:2009wy,Fu:2010ay,Bhattacharyya:2010ef,Bhattacharyya:2014uxa,Shao:2017yzv} 
and at finite temperature and density in 
\cite{Fu:2010ay,Shao:2017yzv,Liu:2017qyc}.

Other models have been employed to study higher-order baryon number 
susceptibilities at finite temperature and density like the Polyakov-loop 
extended quark-meson model \cite{Almasi:2017bhq}, where the influence of 
repulsive vector-interactions on this fluctuations was also analyzed, the 
hybrid quark-meson-nucleon model \cite{Marczenko:2017huu}, or the  SU(3) flavor 
parity-doublet quark-hadron model \cite{Mukherjee:2016nhb} where the 
higher-order baryon number susceptibilities near the chiral and the nuclear 
liquid-gas transitions were investigated.

The eventual location of the CEP can be affected by several conditions such as 
the presence of external magnetic fields or the strangeness and isospin content 
of the medium 
\cite{Costa:2013zca,Costa:2015bza,Rechenberger:2016gim}. 
The study of the CEP location has been undertaken using different versions of 
the NJL and PNJL models. In particular, it was shown that the presence of 
repulsive vector interactions affects strongly the position of the CEP. The 
role played by them were analyzed in detail in 
\cite{Fukushima:2008wg,Costa:2015bza}. 
The calibration of these models at high densities requires the existence of 
experimental data or neutron star observables.
Particularly relevant is the introduction of repulsive interactions, namely the
vector-isoscalar terms, that seems to be necessary to describe $2M_\odot$ 
hybrid stars \cite{Pereira:2016dfg}.

The chiral restoration of strange quarks may play an important role 
inside neutron stars. 
In particular, if this transition occurs at densities that can be found inside 
compact stars, pure quark matter \cite{Pereira:2016dfg}, or, exotic quark 
phases such as the color-flavor-locked (CFL) phase could be realized in their 
interior \cite{Alford:2001zr}. 
Besides, a phase transition could also have an important effect on the 
mean-free path of neutrinos in a protoneutron star as discussed in 
\cite{Gulminelli:2013qr}. The cooling of protoneutron stars during the first 
seconds is essentially driven by the neutrinos that diffuse out of the star. 
A phase transition would give rise to a opalescence like phenomena reducing a 
lot the neutrino mean-free path, and, therefore, allowing for a much larger 
interaction of neutrinos with matter.

In this work, we study the phase diagram using the (2+1)-flavor 
Nambu--Jona-Lasinio model coupled to the Polyakov loop, designated as PNJL 
model, from the point of view of the kurtosis and skewness of net-baryon number 
fluctuations. 
It is expected that in HIC the fireball evolves along isentropes, lines with 
constant entropy per baryon, and, therefore, we analyze how these quantities, 
as well as the velocity of sound, behave along isentropes. 
Our main objective is to identify the similarities and differences of a QCD 
phase diagram which has a CEP or not, namely when a 
sufficiently strong repulsive vector interaction is taken into account.

The model is succinctly reviewed in Sec. \ref{sec:model}, while the results are 
discussed in Sec. \ref{sec:Results}.
Finally we draw our conclusions in Sec. \ref{sec:conclusions}. 

\section{Model and formalism}
\label{sec:model}

The Lagrangian density for the Polyakov extended Nambu--Jona-Lasinio (PNJL) 
model reads
\begin{eqnarray}
{\cal L} &=& {\bar{q}} \left[i\gamma_\mu D^{\mu}-
	{\hat m}_c \right ] q ~+~ {\cal L}_\text{sym}~+~{\cal L}_\text{det}~
  +~{\cal L}_\text{vec} \nonumber\\
&+& \mathcal{U}\left(\Phi,\bar\Phi;T\right),
	\label{Pnjl}
\end{eqnarray}
where the quark field is represented by $q = (u,d,s)^T$ in flavor space, and 
${\hat m}_c= {\rm diag}_f (m_u,m_d,m_s)$ is the corresponding (current) mass 
matrix.
The ${\cal L}_\text{sym}$ and ${\cal L}_\text{det}$ denote, respectively, 
the scalar-pseudoscalar and the 't Hooft six-fermion interactions
\cite{Klevansky:1992qe,Hatsuda:1994pi},
\begin{align}
	{\cal L}_\text{sym}&= G_s \sum_{a=0}^8 \left [({\bar q} \lambda_ a q)^2 + 
	({\bar q} i\gamma_5 \lambda_a q)^2 \right ] \\
	{\cal L}_\text{det}&=-K\left\{{\rm det} \left [{\bar q}(1+\gamma_5)q \right] + 
	{\rm det}\left [{\bar q}(1-\gamma_5)q\right] \right \}.
\end{align}
The vector interaction is given by \cite{Mishustin:2000ss}
\begin{equation} 
{\cal L}_\text{vec} = - G_V \sum_{a=0}^8  
\left[({\bar q} \gamma^\mu \lambda_a q)^2 + 
 ({\bar q} \gamma^\mu \gamma_5 \lambda_a q)^2 \right]. 
\label{p1} 
\end{equation}
The effective gluon field is given by 
$A^\mu = g_{strong} {\cal A}^\mu_a\frac{\lambda_a}{2}$, where
${\cal A}^\mu_a$ represents the SU$_c(3)$ gauge field.
The spatial components are neglected in Polyakov gauge at finite temperature,
i.e., $A^\mu = \delta^{\mu}_{0}A^0 = - i \delta^{\mu}_{4}A^4$. 
The Polyakov loop value is defined as the trace of the Polyakov line,
$ \Phi = \frac 1 {N_c} {\langle\langle \mathcal{P}\exp i\int_{0}^{\beta}d\tau\,
A_4\left(\vec{x},\tau\right)\ \rangle\rangle}_\beta$,
which is the order parameter of the $\Z_3$ 
symmetric/broken phase transition in pure gauge.
For the pure gauge sector we use the following effective potential  \cite{Roessner:2006xn},
\begin{eqnarray}
	& &\frac{\mathcal{U}\left(\Phi,\bar\Phi;T\right)}{T^4}
	= -\frac{a\left(T\right)}{2}\bar\Phi \Phi \nonumber\\
	& &
	+\, b(T)\mbox{ln}\left[1-6\bar\Phi \Phi+4(\bar\Phi^3+ \Phi^3)
	-3(\bar\Phi \Phi)^2\right],
	\label{Ueff}
\end{eqnarray}
where 
$a\left(T\right)=a_0+a_1\left(\frac{T_0}{T}\right)+a_2\left(\frac{T_0}{T}\right)^2$, 
$b(T)=b_3\left(\frac{T_0}{T}\right)^3$. 
Its parametrization values are $a_0 = 3.51$, $a_1 = -2.47$, $a_2 = 15.2$, 
and $b_3 = -1.75$ \cite{Roessner:2006xn}, while the critical temperature is set 
to $T_0=210$ MeV.
The divergent ultraviolet sea quark integrals are regularized by a sharp cutoff 
$\Lambda$ in three-momentum space.
For the NJL model parametrization, we consider:
$\Lambda = 602.3$ MeV, $m_u= m_d=5.5$ MeV, $m_s=140.7$ MeV, 
$G_s \Lambda^2= 1.835$, and $K \Lambda^5=12.36$ \cite{Rehberg:1995kh}.\\

Fluctuations of conserved charges, such as the baryon number, provide vital 
information on the effective degrees of freedom and on critical phenomena.
They behave characteristically in a thermal equilibrium medium.
If there is a CEP in the phase diagram of strongly interacting matter, these 
fluctuations are then expected to provide characteristic signatures that, 
hopefully, can be experimentally observed.
For a static system of infinite size, the fluctuations of baryon number diverge 
at the CEP (second-order phase transition point). 
However, the created medium in HIC experiments has both finite size and 
lifetime that restricts its correlation length and, instead of divergent 
fluctuations, only moderate enhancements are expected.
Fluctuations of conserved charge are characterized by their cumulants or 
susceptibilities.
The present work focuses on the baryon number charge susceptibilities. 
The n$th$-order net-baryon susceptibility is given by
\begin{equation}
 \chi_B^n(T,\mu_B)= \frac{\partial^n\pc{P(T,\mu_B)/T^4}}{\partial(\mu_B/T)^n}.
\end{equation}
Different susceptibility ratios $\chi_B^n(T,\mu_B)/\chi_B^m(T,\mu_B)$
are calculated in order to eliminate the volume dependence, allowing for a 
comparison with experimental observables. 
In this work, we analyze the following ratios
\begin{equation}
	\frac{\chi_B^4(T,\mu_B)}{\chi_B^2(T,\mu_B)}=\kappa\sigma^2,\quad
	\frac{\chi_B^3(T,\mu_B)}{\chi_B^1(T,\mu_B)}=\frac{S_B\sigma^3}{M},
\label{eq:ratios}
\end{equation}
where $M=VT^3\chi_B^1$ is the mean, $\sigma^2=VT^3\chi_B^2$ the variance, $S_B$ 
the skewness, and $\kappa$ is the kurtosis of the net-baryon number 
distribution. 

\section{Results}\label{sec:Results}
\begin{figure*}[t!]
	\centering
	\includegraphics[width=0.8\linewidth]{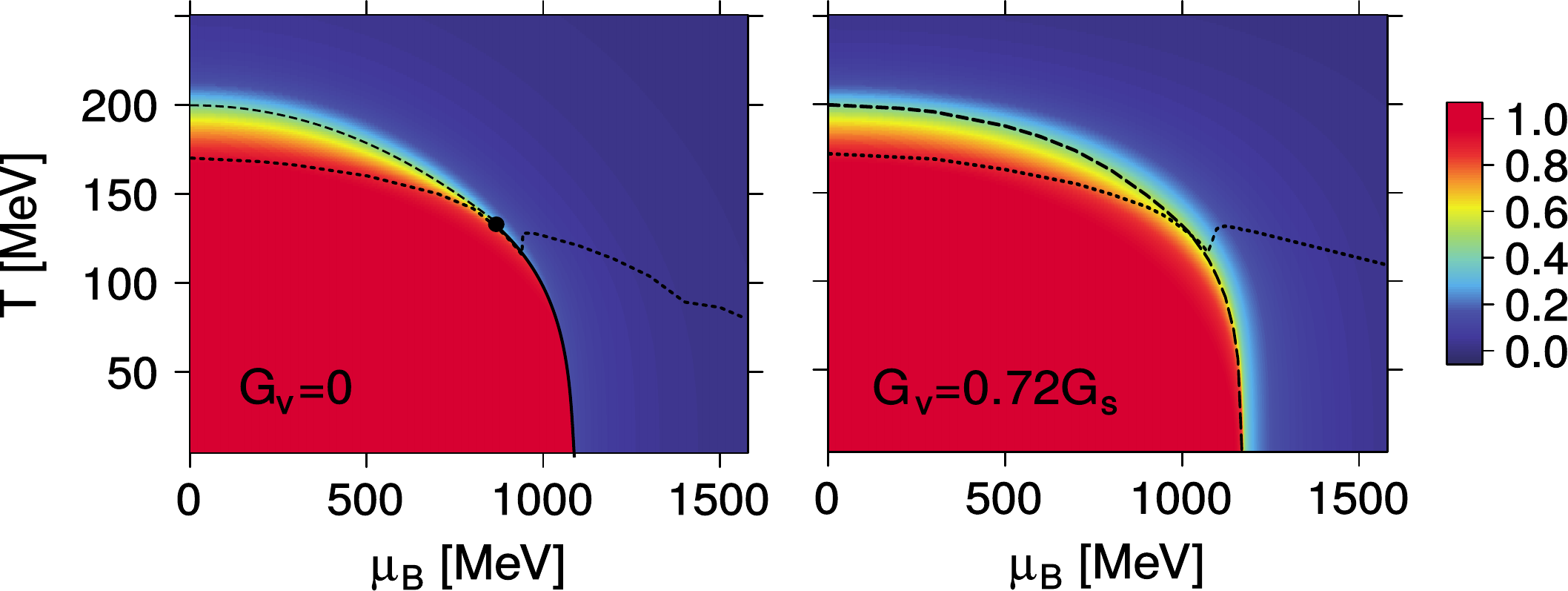}
	\caption{The light-quark condensate 
		$\ev{u\bar{u}}(T,\mu_B)/\ev{u\bar{u}}(0,0)$ for $G_V=0$ (left) and 
		$G_V=0.72G_s$ (right). 
		The following information is displayed: the CEP (dot), the chiral 
		first-order phase transition boundary (solid line), and both the chiral 
		(dashed line) and deconfinement (dotted line) crossover boundaries.} 
	\label{fig:1}
\end{figure*}

We analyze, herein, the net-baryon susceptibilities on the $(T,\mu_B)$ plane.
Two PNJL models are analyzed: (i) a model with no vector interactions $G_V=0$,
which predicts a CEP; 
and (ii) a model with vector interactions $G_V=0.72G_s$, which predicts no CEP.
We want to discuss what distinguishes these two scenarios. 
In the following, symmetric quark matter is considered: 
$\mu_u=\mu_d=\mu_s=\mu_q=\mu_B/3$, where $\mu_{i}$ are the chemical potential of 
each quark flavor and $\mu_B$ is the baryonic chemical potential.

The hydrodynamical expansion of a HIC fireball is expected to follow  
trajectories of constant entropy per baryon, $s/\rho_B$, known as isentropes. 
These trajectories contain important information on the adiabatic evolution of 
the system. 
It is thus interesting to analyze the susceptibility ratios 
[Eqs. (\ref{eq:ratios})] along different isentropes \cite{Costa:2010zw}.
It is important to note that while the net charge and the net strangeness are 
not constrained in the present work; in a HIC, however, the ratio of electric 
charge over baryon number is $Q/\rho_B\simeq0.4$ and no net strangeness is 
produced, $n_s=0$.\\

The phase diagrams for the chiral and deconfinement transitions are presented in 
Fig. \ref{fig:1}. The (normalized) light-quark condensate value
$\ev{u\bar{u}}(T,\mu_B)/\ev{u\bar{u}}(0,0)$ is shown, where  
$\ev{u\bar{u}}(0,0)$ is the vacuum value (due to isospin symmetry 
$\ev{u\bar{u}}=\ev{d\bar{d}}$).
The $G_V=0$ model predicts a CEP at 
$(T^{\mbox{\footnotesize{CEP}}},\mu_B^{\mbox{\footnotesize{CEP}}})=(133\, \text{MeV},862\,\text{MeV})$, 
while the $G_V=0.72G_s$ model has no CEP, and the (approximate) 
chiral restoration is thus attained via an analytic transition (crossover) over 
the whole phase diagram. 
The chiral (dashed line) and deconfinement (dotted line) crossover boundaries 
are determined by the location $(T,\mu_B)$ of the maximum of the order 
parameter susceptibilities (the point where fluctuations are largest).
It is interesting that the crossover boundaries show similar behavior for both 
models:
the gap between the deconfinement and chiral crossovers reduces with increasing 
$\mu_B$ and becomes zero for some $\mu_B$ values, which turns out to be near the 
CEP for $G_V=0$, above which they separate and follow distinct paths.

Both boundaries, the chiral  phase transition and the deconfinement phase 
transition boundaries, are determined from the peaks of the susceptibility. The 
crossing of the deconfinement and  chiral phase transitions has already been 
observed before \cite{Costa:2011fh} and it is possible to identify the crossing 
from the calculation of the susceptibility peaks at fixed temperatures: before 
and after the crossing they are two distinguishable peaks. At the crossing, 
that stretches along a finite range of temperatures,  the two peaks overlap. 
The crossing region includes part of the chiral crossover for both models, and, 
in the case of the model with a CEP, also the CEP, and part of the first-order 
phase transition.

We thus conclude that, due to the mixing between the gluonic and quarkionic 
degrees of freedom, the chiral phase transition has a strong influence on 
deconfinement transition. This is reflected on the behavior of the 
deconfinement transition at the light quark and the strange quark chiral 
transition. At the light quark transition, the crossing temperature is not much 
affected, but the crossing chemical potential is tightly connected with the 
position of the chiral transition and the crossing follows, as referred above,  
the chiral crossover or both the chiral crossover and first-order transition. 
As a consequence, the crossing occurs at a much larger chemical potential for 
the $G_V=0.72 G_s$ model. A similar interconnection is observed at the strange 
chiral crossover in Fig. 2 and 3 in the $G_V=0$ model, where the deconfinement 
transition presents a kink.
	
\begin{figure*}[t!]
	\centering
	\includegraphics[width=0.65\linewidth]{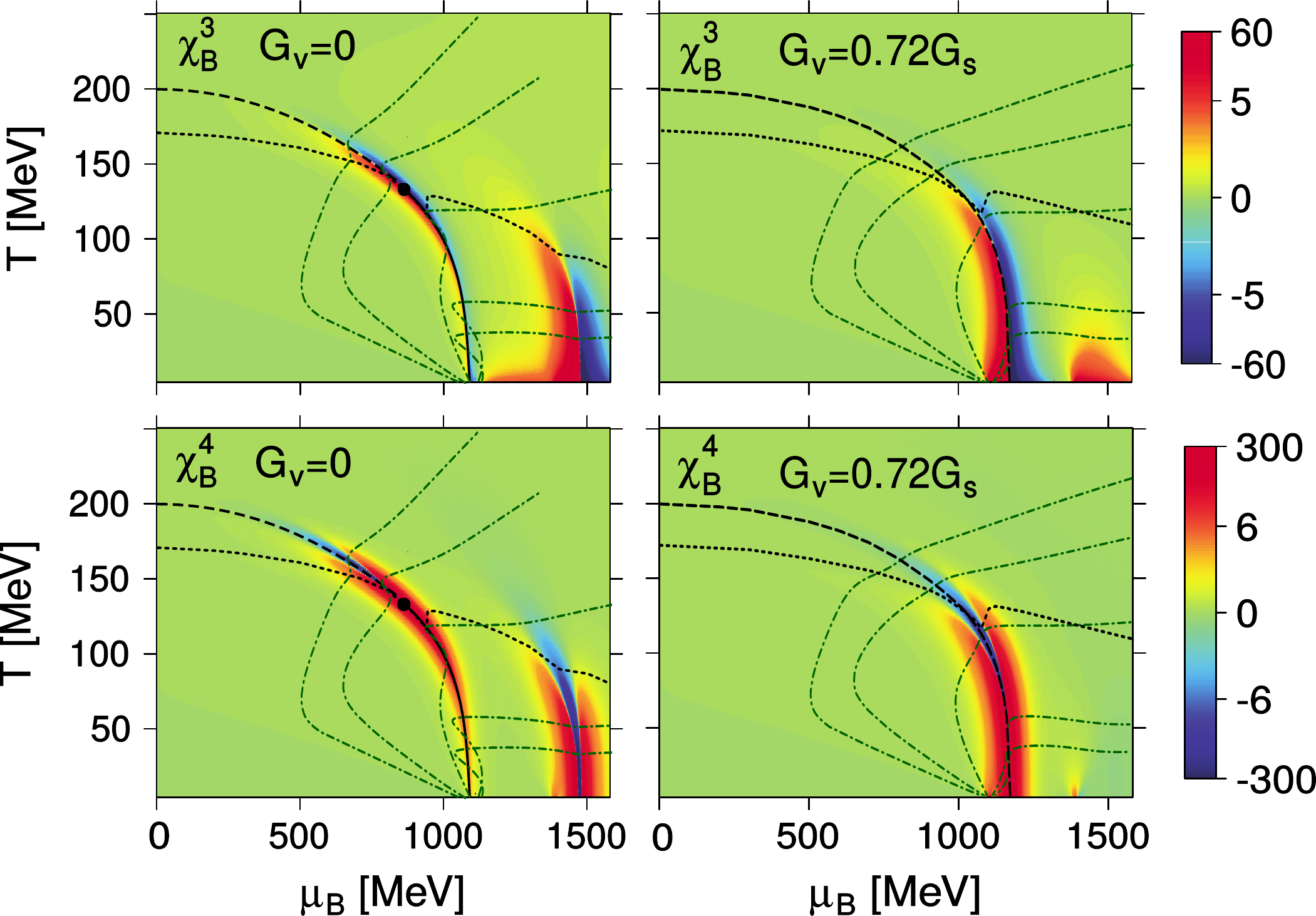}
	\caption{The net-baryon number susceptibilities 
		$\chi^3_B$ (top) and $\chi^4_B$ (bottom) for $G_V=0$ (left) and 
		$G_V=0.72G_s$ (right).
		The following information is displayed: the CEP (black dot), the first-order
		phase chiral transition boundary (black solid line), both the chiral (black 
		dashed line) and deconfinement (black dotted line) crossover boundaries,
		and the $s/\rho_B=\{0.5,1,5,10,14\}$ isentropic trajectories (dark green 
		dotted-dashed lines) are also shown, which appear in the counterclockwise
		direction, respectively.} 
	\label{fig:2}
\end{figure*}
We show  the $\chi_B^3$ (top) and $\chi_B^4$ (bottom) susceptibilities in Fig 
\ref{fig:2}. 
To a better understanding of their dependencies in the $(T,\mu_B)$ plane, 
the following features are also displayed:
the CEP (black dot), the first-order chiral phase transition boundary (black 
solid line), and both the chiral (black dashed line) and the deconfinement 
(black dotted line) crossover boundaries.
Furthermore, the isentropic trajectories (dark green dashed-dotted lines), 
i.e., paths along which the entropy density per baryon, $s/\rho_B$, is fixed, 
are also shown for $s/\rho_B=\{0.5,1,5,10,14\}$. 
The two last trajectories cross the crossover line above the CEP of the $G_V=0$ 
model.

\begin{figure*}[!htbp]
	\centering
	\includegraphics[width=0.65\linewidth]{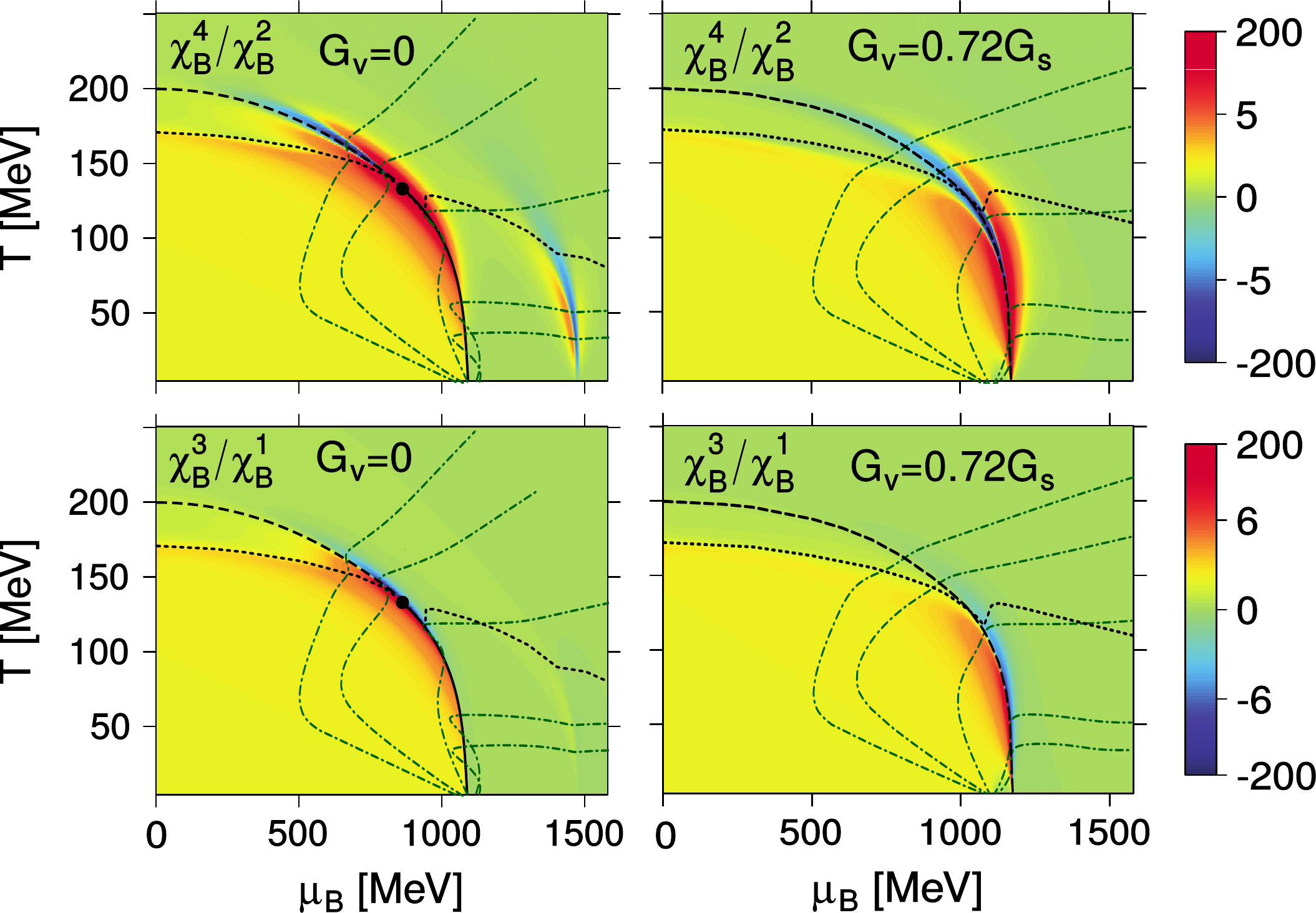}
	\caption{The net-baryon number susceptibility ratios
		$\chi^4_B/\chi^2_B$ (top) and $\chi^3_B/\chi^1_B$ (bottom) 
		for $G_V=0$ (left) and $G_V=0.72G_s$ (right).
		The following information is displayed: the CEP (black dot),
		the first-order phase chiral transition boundary (black solid line), both 
		the chiral (black dashed line) and deconfinement (black dotted line) 
		crossover boundaries, and the $s/\rho_B=\{0.5,1,5,10,14\}$ isentropic 
		trajectories (dark green dotted-dashed lines) are also shown, which appear
		in the counterclockwise direction, respectively.} 
	\label{fig:3}
\end{figure*}

The first three values allows us to discuss the phase diagram at low $T$ and 
high $\mu_B$, where the (approximate) chiral restoration of the strange quark 
occurs.
For the $G_V=0$ model, the susceptibilities exhibit a nonmonotonic dependence 
near the CEP, whose behavior strongly depends on the direction on which the CEP 
is approached. 
The susceptibilities diverge at the CEP, with the divergence being stronger
as higher susceptibilities orders are considered.
An interesting result is present at low $T$ and high $\mu_B$. 
Despite the transition for the strange quark being just a crossover,
and, therefore, without any nonanalytic behavior, a similar CEP 
structure is seen at $\mu_B\approx1500$ MeV for the susceptibilities. 
This indicates that a slight change on the model parametrization might induce 
a first-order phase transition for the strange quark, and a corresponding CEP. 
The $\chi_B^3$ and $\chi_B^4$ values for the $G_V=0.72G_s$ model show precisely 
this behavior for the light quark sector: even though there is no CEP, and the 
chiral transition occurs via a crossover over the whole phase diagram, the 
nonmonotonic behavior of the susceptibilities is still present, as discussed 
within the NJL model \cite{Fan:2017kym}.
The study of a scenario with a hypothetical negative 
$T^{\mbox{\footnotesize{CEP}}}$ for the light CEP, obtained by varying the 
value of the anomaly-induced six-fermion term, $K$, was done in 
\cite{Chen:2016sxn}; it was shown that the magnitude of the susceptibilities 
also changes significantly if a hypothetical negative temperature CEP is taken 
into account.

\begin{figure}[b!]
	\centering
	\includegraphics[width=0.9\linewidth]{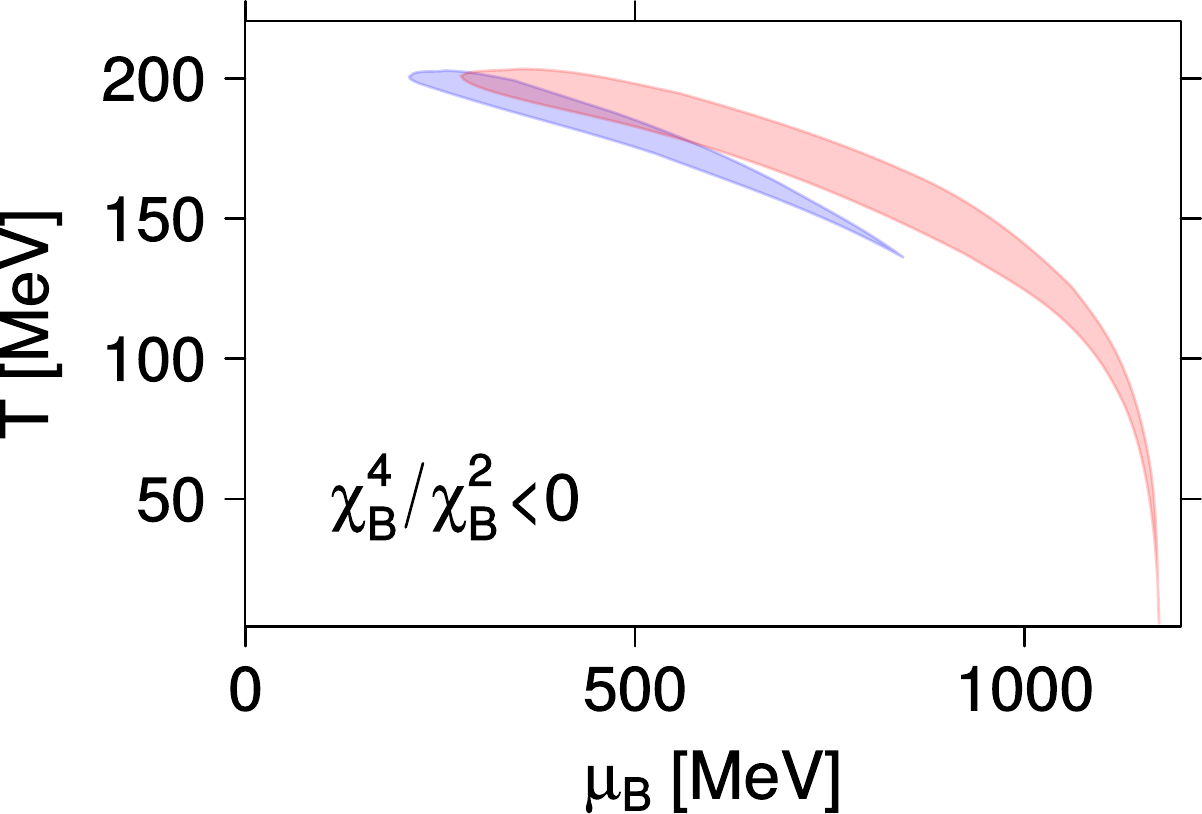}
	\caption{The region $\chi^4_B/\chi^2_B<0$ for $G_V=0$ (blue) and 
	$G_V=0.72G_s$ (red) models.
		} 
	\label{fig:4}
\end{figure}

The ratios $\chi_B^4/\chi_B^2$ and $\chi_B^3/\chi_B^1$ are shown in 
Fig. \ref{fig:3}. The sudden decrease near the deconfinement pseudocritical 
temperature (dotted black line) indicates that both quantities are valuable 
signatures of deconfinement transition.
As noted in \cite{Fu:2009wy}, the statistical confinement, provided by the 
Polyakov loop (at low temperatures, when $\Phi,\bar{\Phi}\rightarrow 0$, 
contributions coming from one- and two-quark states are suppressed, while 
three-quark states are not \cite{Hansen:2006ee}), is essential to obtain a 
low-temperature limit for the susceptibility ratios that is consistent with 
the hadron resonance gas model.
The results for the $G_V=0.72G_s$ model clearly show that the nonmonotonic 
behavior of $\chi_B^3$ and $\chi_B^4$, which signals the presence of a
critical behavior, is still present even in the absence of a CEP. The 
nonmonotonic behavior persists, with a smaller intensity, up to almost the same 
temperature as for the $G_V=0$ model. To make this feature clear, we show the 
negative region of $\chi^4_B/\chi^2_B$ in Fig. \ref{fig:4}. 
Despite the strong vector interaction used, it is remarkable that, for the 
$G_V=0.72G_s$ model (red), the $\chi^4_B/\chi^2_B<0$ region extends from zero 
up to temperatures similar with the ones obtained for $G_V=0$ (blue). 
This indicates that, at higher temperatures, both models are not discernible   
exclusively from the sign change of $\chi^4_B/\chi^2_B$. 
Actually, a region with $\chi^4_B/\chi^2_B<0$ is still present (at lower 
temperatures though) even when the vector interaction strength is increased up 
to $G_V\approx1.4G_s$. 
If instead of looking at the whole negative region of $\chi^4_B/\chi^2_B<0$, 
one considers the stronger fluctuation region $\chi^4_B/\chi^2_B<-200$ the 
following pattern is seen:
while this region extends to a range of $\Delta \mu_B\approx100$ MeV and 
$\Delta T\approx20$ MeV for $G_V=0$, we get $\Delta \mu_B\approx20$ MeV and 
$\Delta T\approx60$ MeV for $G_V=0.72G_s$.  
These different ranges on $T$ and $\mu_B$ for the two different 
scenarios could help distinguish them, taking only into account the behavior of 
the fluctuations. It should, however, be recalled that if the CEP exists, for 
moderate temperatures and high enough baryonic density the line of first-order 
transition could be crossed during the evolution of the fireball, giving rise 
to effects like multifragmentation 
\cite{Mishustin:2006ka,Braun-Munzinger:2015hba}. 
This would be a region where our no CEP model would present fluctuations 
similar to the ones existing in a model with CEP, above the CEP.

\begin{figure}[b!]
	\centering
	\includegraphics[width=0.80\linewidth]{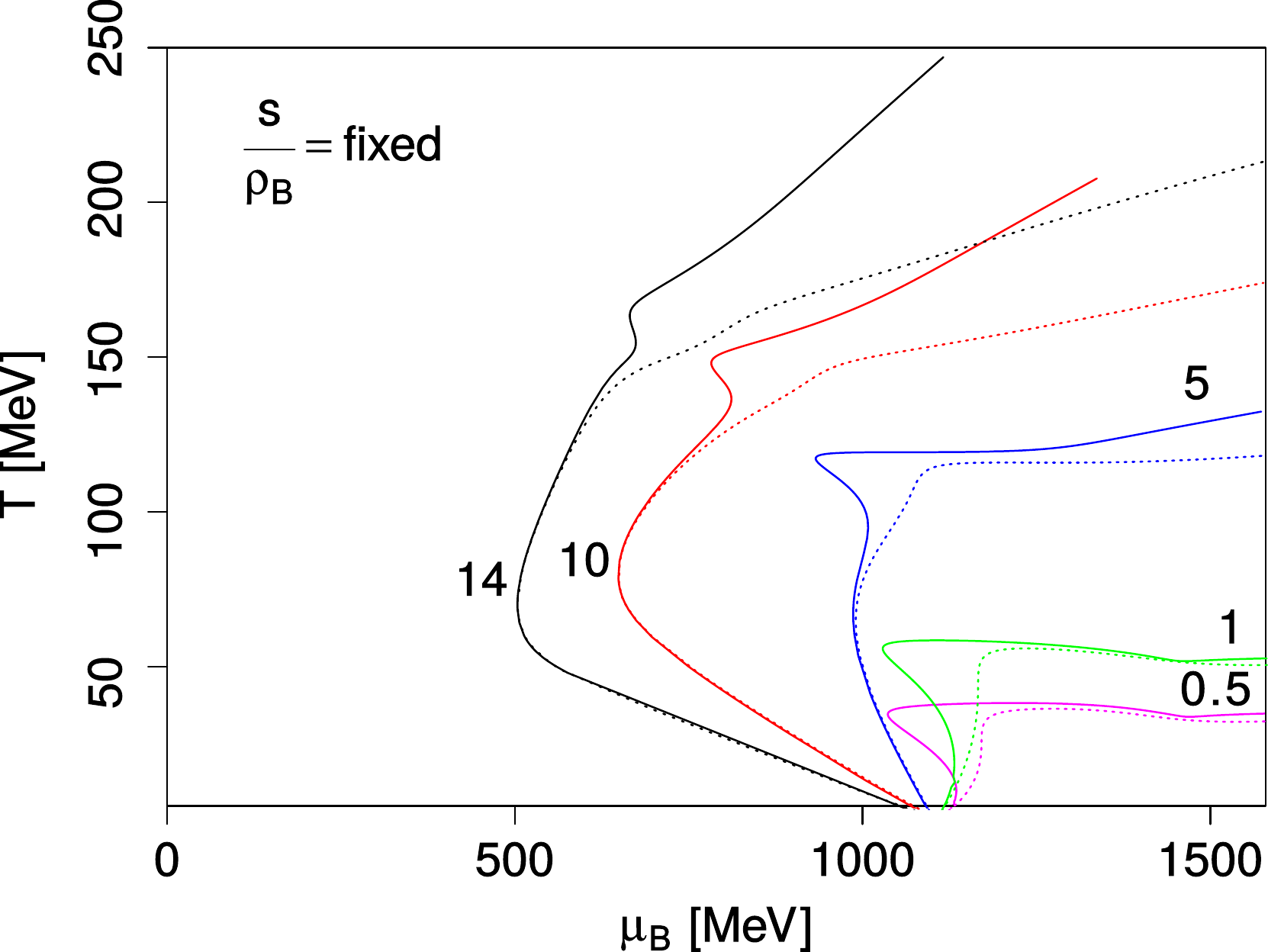}
	\caption{Isentropic trajectories $s/\rho_B=\{0.5,1,5,10,14\}$ for 
	the $G_V=0$ (solid lines) and the $G_V=0.72G_s$ (dotted lines) models.} 
	\label{fig:5}
\end{figure}

To complete the discussion, in the following we analyze the  
isentropic trajectories with a small $s/\rho_B$ within the two scenarios.

The comparison of the isentropic trajectories between $G_V=0$ (solid lines) 
and $G_V=0.72G_s$ (dashed lines) models is in Fig. \ref{fig:5}.  
The trajectories differ for high values of $T$ and $\mu_B$, i.e., as soon as 
the system becomes denser enough for the vector interactions to set in.
Two features that distinguish the $G_V=0$ model from the $G_V=0.72G_s$
is the behavior of the trajectories near the CEP and the existence of a 
unstable spinodal region. 
The trajectories with low $s/\rho_B$ values get enclosed into the unstable 
spinodal region when crossing the first-order phase transition to the chiral 
broken phase. 
As the system enters into the unstable spinodal region, the rapid formation of 
fragments of high density matter that occur should enhance the baryon number 
fluctuations \cite{Braun-Munzinger:2015hba}.
Due to the absence of spinodal region for the $G_V=0.72G_s$ model, such effect does not occur and the susceptibilities have an analytic behavior.

\begin{figure}[!t]
	\centering
	\includegraphics[width=0.98\linewidth]{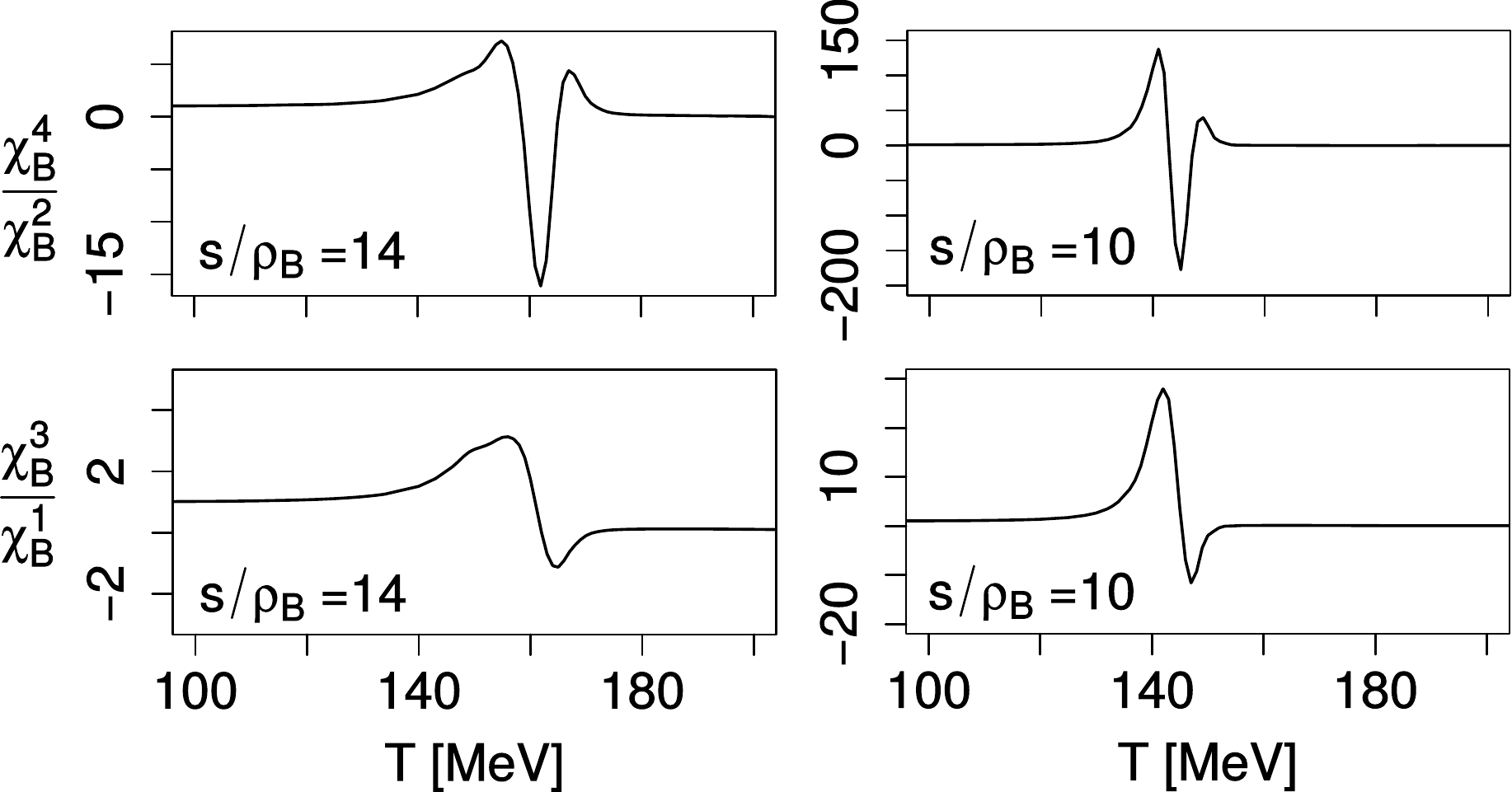}
	\caption{The values of $\chi_B^4/\chi_B^2$ (top)
	and $\chi_B^3/\chi_B^1$ (bottom) as a function of temperature along the isentropes
	$s/\rho_B=14$ (left) and $s/\rho_B=10$ (right) for the $G_V=0$ model.} 
	\label{fig:6}
\end{figure}

In Fig. \ref{fig:6}, we show the $\chi_B^4/\chi_B^2$ (top) and 
$\chi_B^3/\chi_B^1$ (bottom) values along the $s/\rho_B=14$ (left) and the 
$s/\rho_B=10$ (right) isentropes (these isentropic trajectories are shown in 
Fig. \ref{fig:2} and \ref{fig:3}).
As the value $s/\rho_B$ of the isentropic trajectory decreases, we are covering 
a higher $\mu_B$ region on the phase diagram. 
As we move from $s/\rho_B=14$ to $s/\rho_B=10$, we are then approaching a 
region of higher baryon fluctuations that reflects the vicinity of a CEP.
While the fluctuations of $\chi_B^4/\chi_B^2$ and $\chi_B^3/\chi_B^1$
grow with decreasing $s/\rho_B$,
they also become constrained to a smaller temperature region (this is clear 
through the shape of the blue region in Fig. \ref{fig:4}). 
The decreasing gap between the chiral and deconfinement transitions with 
increasing $\mu_B$, which vanishes at the CEP (see Figs. \ref{fig:2} and 
\ref{fig:3}), is also reflected in the fluctuations: for $s/\rho_B\geq14$ a two 
peak structure is present on the left side of the fluctuation (for 
$s/\rho_B=14$, a small bump at $T\approx 150$ MeV is barely seen).

\begin{figure}[!t]
	\centering
	\includegraphics[width=0.98\linewidth]{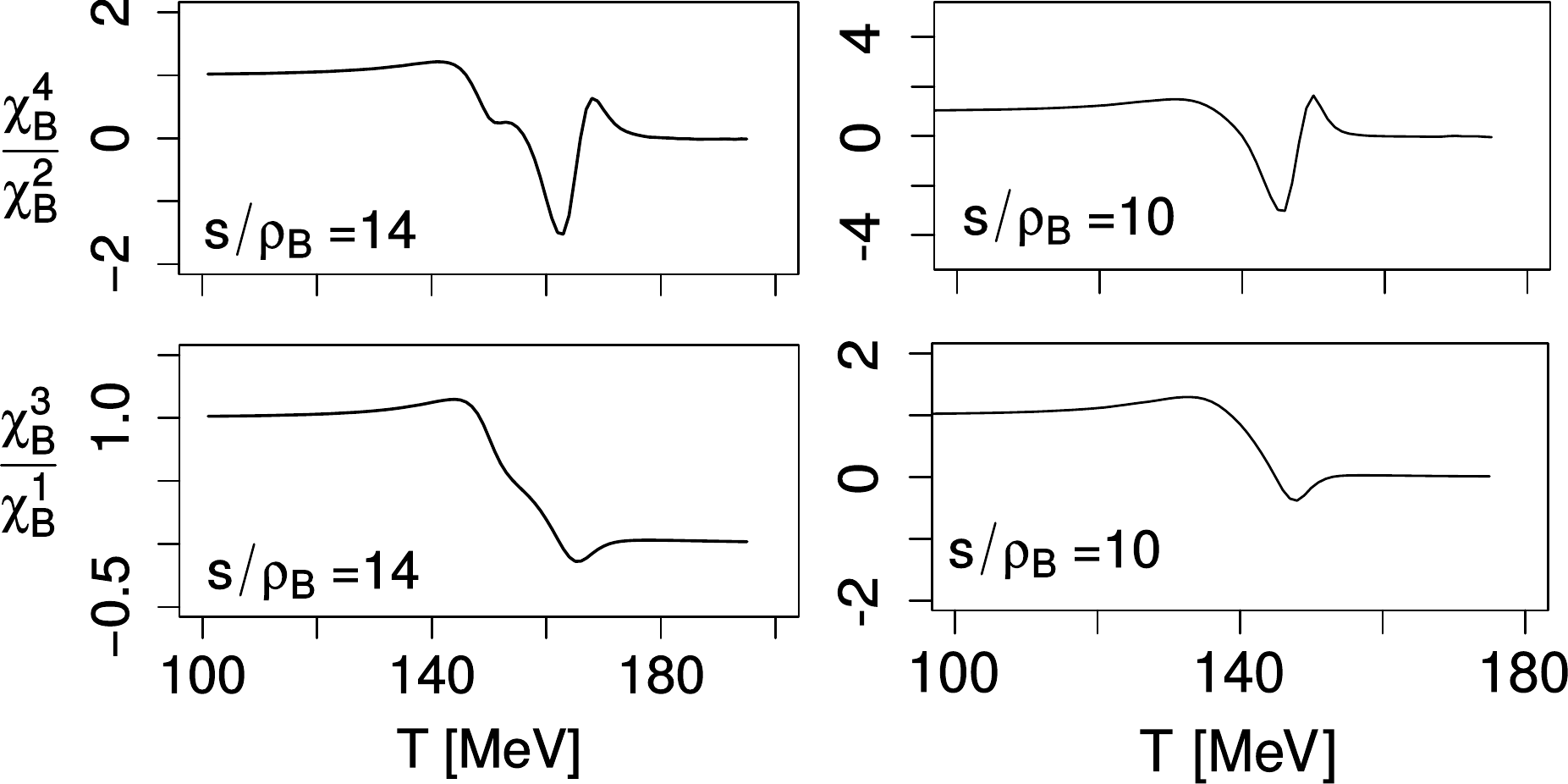}
	\caption{The values of $\chi_B^4/\chi_B^2$ (top)
	and $\chi_B^3/\chi_B^1$ (bottom) as a function of temperature along the isentropes
	$s/\rho_B=14$ (left) and $s/\rho_B=10$ (right) for the $G_V=0.72G_s$ model.} 
	\label{fig:7}
\end{figure}

This two peak structure, which reflects the deconfinement/chiral restoration 
gap, is clearer when a vector interaction is included. 
The fluctuations for the $G_V=0.72G_s$ model, over the same isentropic 
trajectories, are shown in Fig. \ref{fig:7}.
The fluctuations along the $s/\rho_B=14$ trajectory show a two peak structure 
again on the left side of the fluctuation. 
Despite the existence of a sign change of $\chi_B^4/\chi_B^2$ (top) for both 
models (also seen in Fig. \ref{fig:4}), their intensity is weaker for the 
$G_V=0.72G_s$ model, allowing one to notice the effect of the deconfinement
transition on the fluctuations ratios.

\begin{figure}[!b]
	\centering
	\includegraphics[width=0.7\linewidth]{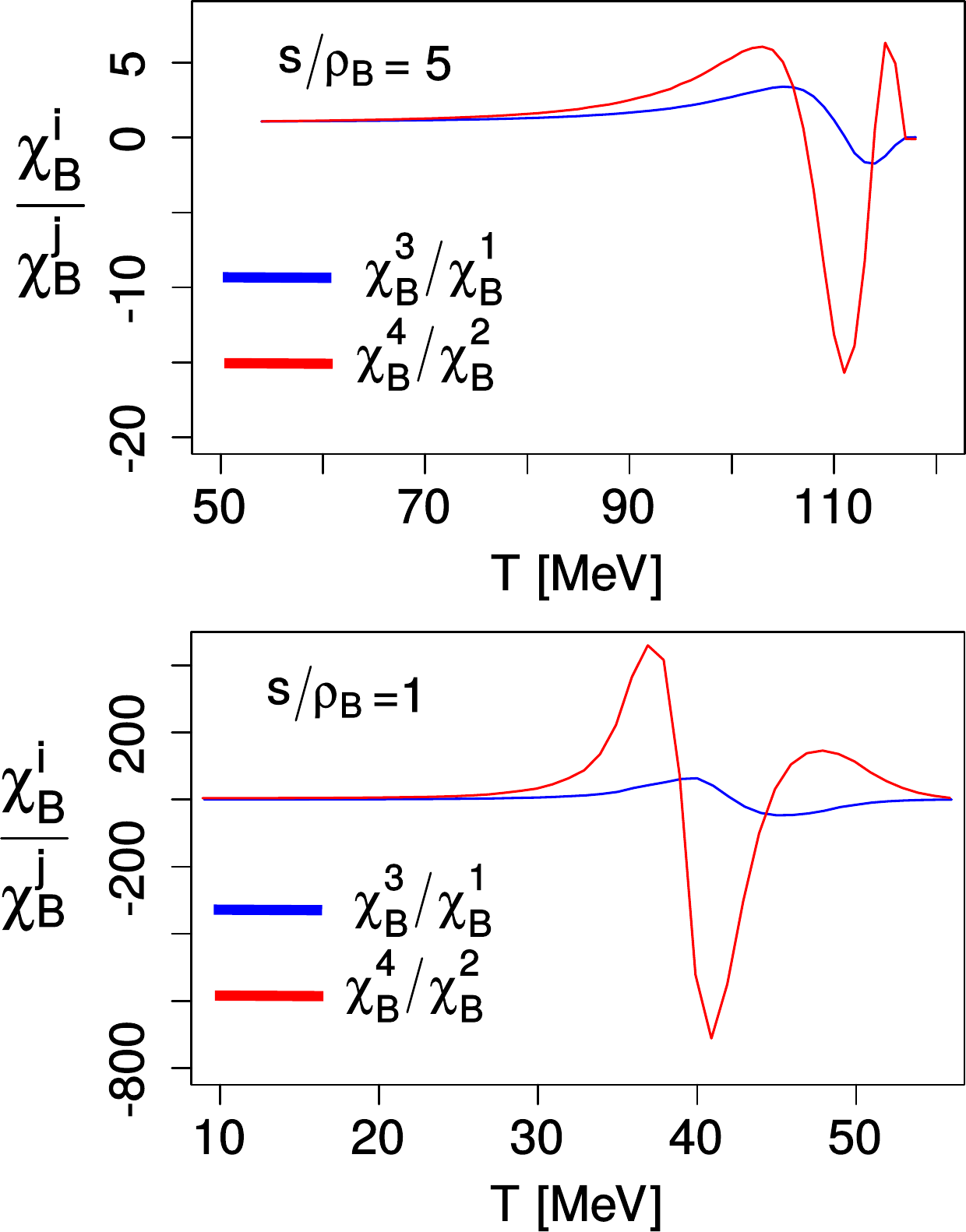}
	\caption{The value of $\chi^4_B/\chi^2_B$ (red) and $\chi^3_B/\chi^1_B$ (blue) 
		as a function of temperature along the isentropic trajectories $s/\rho_B=5$ (left) and $s/\rho_B=1$ 
		(right) for the $G_V=0.72G_s$ model.} 
	\label{fig:8}
\end{figure}

Let us now focus on the crossover region at low temperatures, i.e., low 
$s/\rho_B$ values, for the $G_V=0.72G_s$ model.
In Fig. \ref{fig:8}, we display the values of $\chi^4_B/\chi^2_B$ (red) and 
$\chi^3_B/\chi^1_B$ (blue) along the isentropes $s/\rho_B=5$ (top) and 
$s/\rho_B=1$ (bottom). The $(T,\mu_B)$ dependence of the isentropic 
trajectories can be seen in Figs. \ref{fig:2} and \ref{fig:3}. 
The fluctuations increase strongly as lower isentrope values are considered. 
The large fluctuation of $\chi^4_B/\chi^2_B$ for $s/\rho_B=1$ reflects the 
crossing of the isentropic trajectory with the chiral crossover line at 
$T\approx 40$ MeV.
The features obtained at these low $s/\rho_B$ values are similar with the ones 
of the model with CEP but at $s/\rho_B=10$ and $14$; i.e., we get similar 
fluctuation amplitudes for a much lower $T$.

\begin{figure*}[!tb]
	\centering
	\includegraphics[width=0.65\linewidth]{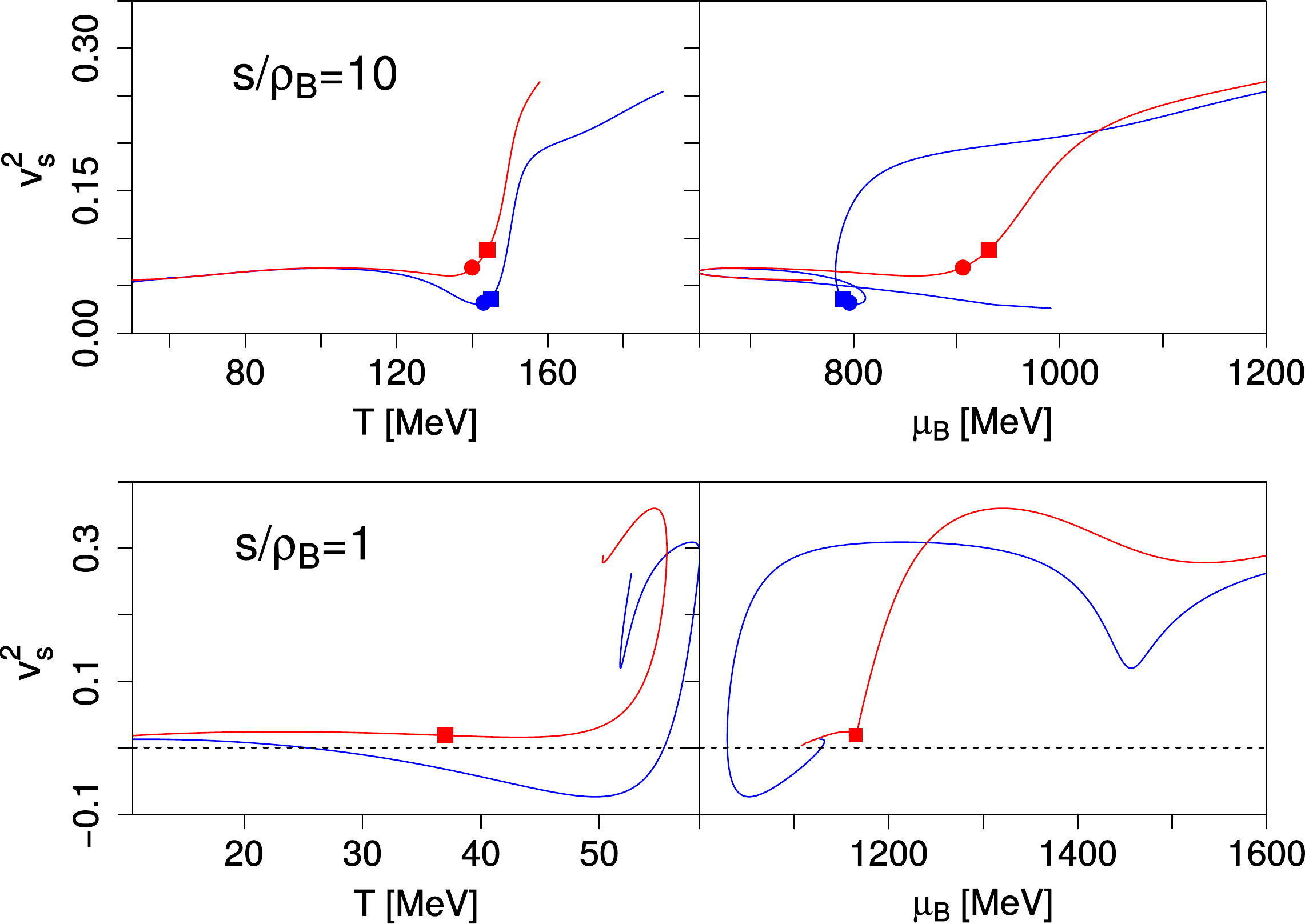}
	\caption{Sound velocity squared $v_s^2$ as a function 
		of temperature (left) and baryon chemical potential (right) along the 
		isentropic trajectories $s/\rho_B=10$ (top) and $s/\rho_B=1$ (bottom) for 
		$G_V=0$ (blue) and $G_V=0.72G_s$ (red) models. 
		The squares and circles indicate the location of
		the chiral and the deconfinement pseudocritical boundaries, respectively.} 
	\label{fig:9}
\end{figure*}

\begin{table*}[t!]
\centering
\begin{tabular}{|c||c|c|c||c|c|c||c|c|c|}
\hline
\multirow{2}{*}{$s/\rho_B$} & $T^{\chi}$ & $\mu_B^{\chi}$ & \multirow{2}{*}{$(v_s^2)^{\chi}$} & $T^{\Phi}$ & $\mu_B^{\Phi}$ & \multirow{2}{*}{$(v_s^2)^{\Phi}$} & $T^{min}$ & $\mu_B^{min}$ & \multirow{2}{*}{$(v_s^2)_{min}$} \\
 & [MeV] & [MeV] & & [MeV] & [MeV] & & [MeV] & [MeV] &  \\
\hline
\multicolumn{10}{|c|}{$G_V=0$ } \\
\hline
8 & 133 & 861 & 0.018 & 133 & 863 & 0.018 & 133 & 863 & 0.018 \\
\hline
10 & 145 & 790 & 0.036 & 143 & 796 & 0.032 & 141 & 803 & 0.031 \\
\hline
12 & 153 & 727 & 0.053 & 148 & 738 & 0.040 & 147 & 738 & 0.039 \\
\hline
14 & 161 & 667 & 0.070 & 153 & 672 & 0.044 & 150 & 668 & 0.042 \\
\hline
20 & 176 & 525 & 0.10 & 161 & 504 & 0.045 & 150 & 668 & 0.042 \\
\hline
\multicolumn{10}{|c|}{$G_V=0.72G_s$}  \\
\hline
0.1 & 5 & 1172 & 0.0049 & $-$ & $-$ & $-$ & 6 & 1173 & 0.0016 \\
\hline
1 & 37 & 1165 & 0.019 & $-$ & $-$ & $-$ & 43 & 1166 & 0.016 \\
\hline
10 & 144 & 931 & 0.088 & 140 & 906 & 0.069 & 134 & 859 & 0.060 \\
\hline
12 & 153 & 870 & 0.10 & 147 & 819 & 0.067 & 141 & 768 & 0.059 \\
\hline
14 & 161 & 814 & 0.11 & 152 & 737 & 0.064 & 174 & 689 & 0.056 \\
\hline
20 & 185 & 554 & 0.18 & 162 & 512 & 0.048 & 159 & 494 & 0.042 \\
\hline\end{tabular}
\caption{The sound velocity squared at the chiral 
		$(v_s^2)^{\chi}=(T^{\chi},\mu_B^{\chi})$ and deconfinement 
		$(v_s^2)^{\Phi}=(T^{\Phi},\mu_B^{\Phi})$ pseudocritical boundaries for
		several isentropes $s/\rho_B$. The minimum of $v_s^2$ [$(v_s^2)_{min}$] and
		its location $(T^{min},\mu_B^{min})$ is also presented. }
\label{Table1}
\end{table*}

Finally, we have determined the square of the sound velocity, 
$v_s^2=dP/d{\cal E}|_{s/\rho_B=\mbox{const.}}$, along two isentropic trajectories. 
The sound velocity plays a central role in the hydrodynamical evolution of 
matter created in HIC being very different in the different stages of the 
expansion. It affects, among others, the momentum distribution of the particles 
originating from the fluid elements at the freeze-out stage 
\cite{Mohanty:2003va}. 
The values of square sound velocity are extracted from the widths of rapidity 
distributions \cite{Mohanty:2003va,Gao:2015sdb,Adam:2016ddh}. For example, from 
the measured data on the widths of the pion rapidity spectra, $v_s^2$ in the 
dense stage of the reactions has been extracted \cite{Steinheimer:2012bp}.

In Fig. \ref{fig:9}, we show $v_s^2$ along $s/\rho_B=10$ (top) and 
$s/\rho_B=1$ (bottom) for $G_V=0$ (blue) and $G_V=0.72G_s$ (red) models.
As the isentropic trajectories follow specific paths, $(T,\mu_B)$, on the phase 
diagram (see Figs. \ref{fig:2} and \ref{fig:3}), we show the $v_s^2$ dependence 
on temperature (left) and baryon chemical potential (right), for both 
isentropic trajectories.
For each isentropic, we give the values of $v_s^2(T,\mu_B)$ at its minimum and 
at the chiral and deconfinement boundaries in Table \ref{Table1}. 

Considering in first place the $s/\rho_B=10$ trajectory as a function of 
$T$ (Fig. \ref{fig:9}, left upper panel), we see that the minimum of 
$v_s^2(T,\mu_B)$ occurs closer to the deconfinement pseudocritical boundary 
(circles) than to the chiral pseudocritical boundary (squares).
While the temperature dependence of the $s/\rho_B=10$ isentrope is a 
single-valued function, the same does not hold for its $\mu_B$ dependence 
(Fig. \ref{fig:9}, right upper panel).
The loop behavior for the $G_V=0$ model (blue curve in upper right panel of 
Fig. \ref{fig:9}) 
rises from the bending effect towards the CEP that the $s/\rho_B=10$ isentrope 
undergoes when crossing into the chiral broken region (solid red line in Fig. 
\ref{fig:5}). 
This effect occurring in $v_s^2$ can then be seen as a signal for the vicinity 
of a CEP (if some kind of bending effect into the CEP exists), once this effect 
is not seen for the $G_V=0.72G_s$ model.
For $s/\rho_B=1$ (lower panels of Fig. \ref{fig:9}), the $v_s^2$ shows negative 
values for $G_V=0$ model (blue curve), reflecting the first-order phase
transition that occurs at lower $T$.
It is interesting to note that, for small values of $s/\rho_B$, the local 
minimum of $v_s^2$ at $\mu_B\approx1480$ MeV is associated  with the crossover
of the strange quark.

\section{Conclusions}\label{sec:conclusions}

We have analyzed the net-baryon number fluctuations for three-flavor quark 
matter within the Polyakov extended Nambu--Jona-Lasinio model.
For a strong enough vector interaction intensity, the model predicts no CEP
in the phase diagram.
From  the net-baryon number fluctuations one concludes that, even in the 
absence of a CEP, the nonmonotonic behavior persists.
Therefore, the existence of a CEP cannot be taken solely from the existence of 
nonmonotonic behavior on the net-baryon number susceptibilities.

We have analyzed further other possible properties that may
distinguish the two scenarios: for the no CEP model ($G_V=0.72G_s$),
large fluctuations in the susceptibility ratios occur only at small $T$, and 
the values of $v_s^2$ are almost unchanged at moderates $s/\rho_B$ values.
The values of the susceptibility ratios along two or three isentropic lines 
would possibly allow us to distinguish both cases. Also, the value of the sound 
velocity at the chiral transition for two or three isentropes would give some 
useful information. 
For the $G_V=0$ model, by going from $s/\rho_B=9$ to $14$, the value of $v_s^2$ 
increases at least $50\%$ for each step $\Delta(s/\rho_B)=2$. 
Instead, for the no CEP model, the change is of the order of $10\%$.
It should be noticed, however, that in the present work we have discussed 
infinite size matter. For a finite system it is expected that the signals we 
have discussed are less intense but still might allow to distinguish both 
scenarios. 

We have shown that, for high chemical potentials and low temperatures,
a signature of the strange quark chiral symmetry restoration is observed 
in a decrease of the sound velocity and in a region with negative 
$\chi^4_B/\chi^2_B$. 

\vspace{0.25cm}
{\bf Acknowledgments}
This work was supported by ``Fundação para a Ciência e Tecnologia,'' Portugal, 
under Projects No. UID/FIS/04564/2016 and No. POCI-01-0145-FEDER-029912 with
financial support from POCI, in its FEDER component, and by the FCT/MCTES 
budget through national funds (OE), and under Grants No. SFRH/BPD/102273/2014 
(P.C.) and No. CENTRO-01-0145-FEDER-000014 (M.F.) through the CENTRO2020 
program. Partial support comes from ``THOR'' (COST Action CA15213) and 
``PHAROS'' (COST Action CA16214).

\vspace{-0.5cm}
\appendix
\section{FORMALISM}
The thermodynamical potential for the three-flavor PNJL model reads	
\begin{align*}
\Omega(T, \mu_i) &= G_{_{s}}\sum_{i=u,d,s}\ev{\bar{q_{i}}q_{i}}^2 
+ 4 K\ev{\bar{q_{u}}q_{u}}\ev{\bar{q_{d}}q_{d}}\ev{\bar{q_{s}}q_{s}} \\
& +G_V \left( \rho_u + \rho_d + \rho_s \right)^2 + {\cal U}\left(\Phi,\bar{\Phi},T\right)\\
& - 2 N_c\sum_{i=u,d,s}\int\frac{\mathrm{d}^3p}{\left(2\pi\right)^3}\chav{{E_i} \theta(\Lambda^2-\mathbf{p}^{2})} \\
& - 2 N_c\sum_{i=u,d,s}\int\frac{\mathrm{d}^3p}{\left(2\pi\right)^3}\chav{T\left(z^+_\Phi(E_i) + z^-_\Phi(E_i) \right) } ,
\end{align*}
where  $E_{i}=\sqrt{\mathbf{p}^{2}+M_{i}^{2}}$ is the quasiparticle 
energy of the quark $i$, and $z^{\pm}_\Phi$ represent the following partition 
function densities,
\begin{align}
z^+_\Phi(E_i) &= \ln\left\{ 1 + 3\left( \bar\Phi + \Phi \expp \right) \expp + \expppp \right\} \\
z^-_\Phi(E_i) &= \ln\left\{ 1 + 3\left(\Phi + \bar\Phi \expm \right) \expm + \expmmm \right\} ,
\end{align}
where $E_i^{(\pm)}\,=\,E_i\,\mp \tilde{\mu}_i$ with the upper (lower) sign 
applying for fermions (antifermions), and $\beta=1/T$.
The quark effective chemical potentials are given by
$$  
\tilde{\mu}_i=\mu_i-4G_V\rho_i
$$
The $i$ quark number density is determined by $\rho_i = - (\partial\Omega/\partial\mu_i)$ and reads
\begin{equation}
\rho_i=2N_c\int \frac{d^3p}{(2\pi)^3}\left(f^{(+)}_\Phi(E_i)-f^{(-)}_\Phi(E_i) \right).
\end{equation}
The modified Fermi-Dirac distribution functions $f^{(+)}_\Phi(E_i)$ and 
$f^{(-)}_\Phi(E_i)$ are given by
\begin{align}
f^{(+)}_\Phi(E_i) & = \frac{ \bar\Phi\expp + 2\Phi\exppp + \expppp }
{ 1+ 3\left( \bar\Phi + \Phi \expp \right) \expp+ \expppp } \\
f^{(-)}_\Phi(E_i) & = \frac{ \Phi\expm + 2\bar\Phi\expmm + \expmmm }
{ 1 + 3\left(\Phi + \bar\Phi \expm \right) \expm+ \expmmm } \label{fpPhi}.
\end{align}
In the mean-field approximation, the values of the quark condensates are given by 
\begin{equation}
\left\langle\bar{q}_{i}q_{i}\right\rangle\,=\,-\,\,2N_c\int\frac{\mathrm{d}^3p}{\left(2\pi\right)^3}
\frac{M_i}{E_i}[\theta(\Lambda^2-\mathbf{p}^{2})-f^{(+)}_\Phi(E_i)-f^{(-)}_\Phi(E_i)],
\end{equation}
which satisfy the following gap equations:
\begin{align}
M_u&=m_u-2G_s\ave{\bar{q}_uq_u}-2K\ave{\bar{q}_dq_d}\ave{\bar{q}_sq_s}\\
M_d&=m_d-2G_s\ave{\bar{q}_dq_d}-2K\ave{\bar{q}_sq_s}\ave{\bar{q}_uq_u}\\
M_s&=m_s-2G_s\ave{\bar{q}_sq_s}-2K\ave{\bar{q}_uq_u}\ave{\bar{q}_dq_d}.
\end{align}
The values of $\Phi$ and $\bar{\Phi}$ are the solutions of 
\begin{align}
0 &= {T^4} \left\{ -\frac{a(T)}{2}\bar\Phi - 6\frac{b(T)\left[\bar\Phi
	-2\Phi^2+\bar\Phi^2 \Phi\right]}
{1-6 \bar\Phi \Phi + 4(\bar\Phi^3 + \Phi^3)-3(\bar\Phi \Phi)^2}\right\} \nonumber\\
&- 6 T \sum_{i=u,d,s}
\int\frac{\mathrm{d}^3p}{\left(2\pi\right)^3}
\left( \frac {\exppp}{ \exp\{z^+_\Phi(E_i)\} } + \frac {\expm}{\exp\{ z^-_\Phi(E_i)\} } \right)
\label{eq:domegadfi} 
\end{align}
\begin{align}
0 &= {T^4} \left\{ -\frac{a(T)}{2}\Phi - 6\frac{b(T)\left[\Phi -2\bar\Phi^2+\bar\Phi
	\Phi^2\right]} {1-6 \bar\Phi \Phi + 4(\bar\Phi^3 + \Phi^3)-3(\bar\Phi \Phi)^2}\right\}
\nonumber\\
&- 6 T \sum_{i=u,d,s}
\int\frac{\mathrm{d}^3p}{\left(2\pi\right)^3}
\left( \frac {\expp}{ \exp\{z^+_\Phi(E_i)\} } +
\frac {\expmm}{\exp\{ z^-_\Phi(E_i)\} } \right).
\label{eq:domegadfib} 
\end{align}
The thermodynamical quantities are determined via the thermodynamical potential 
(see \cite{Costa:2010zw}).
The pressure is given $P(T,\mu_i)=-\Omega(T,\mu_i)$, the density of the 
$i$ quark, $\rho_i$, is given by
\begin{equation}\label{dens}
\rho(T,\mu_i)\,= \left. \frac{\partial p}{\partial {\mu_i}}\right|_{T}\,,
\end{equation}
while the the entropy density, $s$, is given by
\begin{equation}\label{entropy}
s(T,\mu_i)\,= \left. \frac{\partial p}{\partial T}\right|_{\mu_i}\,.
\end{equation}
The energy density, ${\cal E}$, comes from the following fundamental relation of
thermodynamics
\begin{equation}\label{energydens}
    {\cal E} (T, \mu_i)\,=\,T\,s(T,\mu_i)-\,p(T,\mu_i)+\sum_{i=u,d,s}\mu_i\rho_i\,.
\end{equation}


\end{document}